\documentclass[aps,pre,reprint,doi=false,isbn=false,url=false,title=true,longbibliography,noeprint,footinbib,onecolumn]{revtex4-1}
\usepackage{graphicx}
\usepackage{amsmath}
\usepackage{array}
\usepackage{amsfonts}
\usepackage{float}
\usepackage{xcolor}
\usepackage[mathscr]{euscript}
\def\ba{\begin{eqnarray}}
\def\ea{\end{eqnarray}}
\def\be{\begin{equation}}
\def\ee{\end{equation}}
\def\bm{\begin{math}}
\def\me{\end{math}}

\newcommand{\dummy}

\makeatletter
\newcommand{\fmarki}{*}
\newcommand{\fmarkii}{\ensuremath{\dagger}}
\newcommand{\fmarkiii}{\ensuremath{\ddagger}}
\newcommand{\fmarkiv}{\ensuremath{\mathsection}}
\newcommand{\fmarkv}{\ensuremath{\mathparagraph}}
\newcommand{\fmarkvi}{\ensuremath{\|}}
\newcommand{\fmarkvii}{**}
\newcommand{\fmarkviii}{\ensuremath{\dagger\dagger}}
\newcommand{\fmarkix}{\ensuremath{\ddagger\ddagger}}
                
\def\@fnsymbol#1{{\ifcase#1\or \fmarki\or \fmarkii\or \fmarkiii\or \fmarkiv\or \fmarkv\or \fmarkvi\or \fmarkvii\or \fmarkviii\or \fmarkix \else\@ctrerr\fi}}
\makeatother

\renewcommand{\fmarki}{$\dagger$}
 \renewcommand{\fmarkii}{*}
 \renewcommand{\fmarkiii}{$\ddagger$}
 \renewcommand{\fmarkiv}{a$_4$}
 \renewcommand{\fmarkv}{x$_5$}

\begin{document}

\title{Segregation disrupts the Arrhenius behavior of an isomerization reaction}
\author{Shubham Thwal}\email[]{ shubhamthwal1.@gmail.com}
\affiliation{Amity Institute of Applied Sciences, Amity University Uttar Pradesh, Noida 201313,
India
}
\author{Suman Majumder}\email[]{smajumder@amity.edu,suman.jdv@gmail.com}
\affiliation{Amity Institute of Applied Sciences, Amity University Uttar Pradesh, Noida 201313,
India
}

\date{\today}

\begin{abstract}
Co-existence of phase segregation and \emph{interconversion} or \emph{isomerization} reaction among molecular species leads to fascinating structure formation in biological and chemical world. Using Monte Carlo simulations of the prototype Ising model, we explore the chemical kinetics of such a system consisting of a binary mixture of \emph{isomers}. Our results reveal that even though the two concerned processes are individually Arrhenius in nature, the Arrhenius behavior of the \emph{isomerization} reaction gets significantly disrupted due to an interplay of the nonconserved dynamics of the reaction and the conserved diffusive dynamics of phase segregation. The approach used here can be potentially adapted to understand reaction kinetics of more complex reactions.
\end{abstract}

% \pacs{82.35.Lr, 64.70.km, 87.15.A-}

\maketitle
The phenomenon of existence of two or more molecular species having the same chemical formula but different properties is referred to as \emph{isomerism} \cite{mcnaught1997}. Structural chirality is one of the reasons behind such \emph{isomerism}, giving rise to \emph{optical isomers} or \emph{enantiomers} \cite{mcnaught1997,fox2004organic}. For instance, L-Glucose, unlike its \emph{enantiomer} D-glucose, is not an energy source for living organism, as it cannot be phosphorylated during \emph{glycolysis}. The final product of synthesis of such molecular species is often comprised of a mixture of its \emph{enantiomers}. Depending on the kind of interactions among themselves the \emph{enantiomeric} components of such mixtures can spontaneously or inductively segregate from each other \cite{pirkle1981,shieh1994}. Simultaneously, either naturally or owing to an external drive, the \emph{enantiomers} may undergo an \emph{isomerization} or \emph{interconversion} reaction leading to an \emph{enantio}-selective production or phase amplification of one of them \cite{soai1995,shibata1996,shibata1996_jacs,tran1996reaction,Qui1997,ohta1998,viedma2005,lombardo2009}. Such \emph{enantio}-selective processes are ubiquitous in nature as well, e.g., amino acid residues of naturally occurring proteins are mostly L-\emph{enantiomers} \cite{lehninger2005}. In the light of the above discussion, it is crucial to have a microscopic understanding by unravelling the physical laws governing such a phenomenon of segregation of molecular species undergoing \emph{isomerization} reaction \cite{glotzer1994,glotzer1995,carati1997,tong2002,krishnan2015,lamorgese2016,anisimov2018,shumovskyi2021,uralcan2021,petsev2021,bauermann2022}.
\par
Segregation in \emph{enantiomeric} mixtures can be easily understood by simply translating the concepts of kinetics of phase segregation, which has been extensively studied in the past \cite{binder1974,puri_book}, and recently developed further in more complex and realistic scenarios \cite{hyman2014,basu2017,mueller2022}. Similarly, the \emph{interconversion} reaction among  \emph{isomers} can be captured under the essence of \emph{phase ordering} of ferromagnets \cite{bray2002}. In \emph{phase ordering} typically one ends up in a state where majority of the magnetic dipoles point in the same direction, following a quench from high-temperature disordered state to a temperature below the Curie point. Both kinetics of phase segregation in a binary mixture  and \emph{phase ordering} can essentially be modeled by using simple lattice models, e.g., the nearest neighbor Ising model. Although both adaptations of these models produce equivalent thermodynamics, fundamentally their dynamics are different. Combining the two approaches results in an appropriate model for exploring the effect of \emph{isomerization} reaction during \emph{enantiomeric} phase segregation in solids using state of the art Monte Carlo (MC) simulations \cite{glotzer1994,puri1994,glotzer1995,shumovskyi2021}. Recent interests in this regard have shifted toward modelling reactions in solutions using molecular dynamics (MD) simulations \cite{longo2022}, and forceful \emph{interconversion} reaction that preserves the respective initial composition of the \emph{enantiomeric} components \cite{longo2023}. These attempts have successfully explored novel mesoscopic steady-state structures mimicking microphase segregation observed in chemical and biological world. However, answers to some of the fundamental questions, viz., effect of segregation on the reaction kinetics, are still unexplored.
\par
In this letter, we study the  kinetics of an \emph{isomerization} reaction of the following type
\begin{equation}\label{reaction}
 A_1 \rightleftharpoons A_2, 
\end{equation}
where the two \emph{isomers} $A_1$ and $A_2$ are also undergoing segregation from each other. If there is no segregation, the rate constant $k$ of such a simple \emph{isomerization} reaction, as a function of temperature $T$, obeys the Arrhenius behavior given as 
\begin{equation}\label{Arhenius_eqn}
 \ln k= \ln A - \frac{E_a}{R}\left(\frac{1}{T}\right),
\end{equation}
where $A$ is a pre-exponential constant, $E_a$ is the activation energy, and $R$ is the universal gas constant. Our results from MC simulations of the prototype Ising model mimicking the system described above, reveal that at high reaction probability the Arrhenius behavior is maintained. However, as the reaction probability decreases and segregation dominates, a significant deviation from the Arrhenius behavior is observed. 
\par
We have chosen a square lattice model where on each site $i$ there sits an Ising spin $S_i=+1(\rm{or}~-1)$ that corresponds to species $A_1 (\rm{or}~A_2)$. The interaction energy between the spins are given by the conventional Ising Hamiltonian

\begin{equation}\label{Ising}
 \mathscr{H}=-J\sum_{\langle ij \rangle}S_iS_j,
\end{equation}
where $\langle ij \rangle$ indicates  that only nearest neighbors can interact with each other and $J$ is the corresponding interaction strength. We apply periodic boundary conditions in all possible directions to eradicate any surface effects. The model exhibits an order-disorder transition with a critical temperature $T_c=[2/\ln(1+\sqrt{2})]J/k_B$, where $k_B$ is the Boltzmann constant \cite{onsager1944}. From now onward the unit of temperature is $J/k_B$, and for convenience we have set $J=k_B=1$. In order to capture the essence of a segregating mixture of \emph{isomers} undergoing \emph{isomerization} reaction, we have introduced both Kawasaki spin-exchange dynamics and Glauber spin-flip dynamics \cite{glauber1963,kawasaki1966,barkema_book,landau_book}. In Kawasaki exchange, interchange of positions between a randomly chosen pair of
nearest-neighbor spins is attempted,  facilitating segregation of species. Such a MC move replicates atomic diffusion, and the resultant dynamics is conserved as it keeps individual compositions of the species unaltered. On other hand, in a Glauber spin-flip move, an attempt is made to flip a randomly chosen spin, thus mimicking the \emph{interconversion} or \emph{isomerization} reaction. We consider the forward and backward reactions in \eqref{reaction} to be equally likely. The spin-flip move is nonconserved as it changes the individual composition of the species. Both moves are accepted according to the  standard Metropolis criterion \cite{barkema_book,landau_book}. We start with a \emph{racemic} mixture of the \emph{isomers}, i.e., equal proportion of $A_1$ and $A_2$ are uniformly distributed on the lattice, and then in the simulation we set the temperature to $T<T_c$. At each MC step, the Glauber move is attempted with a  probability $p_r$, while the Kawasaki exchange attempt is executed with a probability $1-p_r$. We choose one MC sweep (MCS) as the unit of time, which refers to $L^2$ attempted MC moves. We perform all our simulations on a square lattice of linear size $L=32$ having $L^2=1024$ \emph{isomers}, at different $T$ for a range of $p_r\in [10^{-4},5\times 10^{-1}]$.
\par
\begin{figure}[t!]
\centering
\includegraphics*[width=0.45\textwidth]{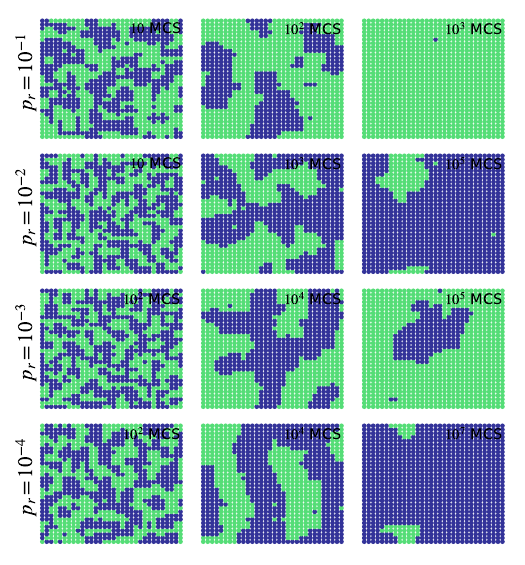}
\caption{\label{snapshots} {\bf Pattern formation due to isomerization reaction and segregation}. Typical snapshots depicting time evolution of a binary mixture of \emph{isomers}, simultaneously undergoing  \emph{isomerization} reaction and segregation, following a quench from a homogeneous phase above $T_c$ to  a temperature $T=0.6T_c$. The results are obtained from simulations on a square lattice of linear size $L=32$. Different rows are for different values of the reaction probability $p_r$, as indicated. Contrasting colors correspond to different species.}
\end{figure}
Segregation of molecular species in combination with the \emph{isomerization} reaction leads to pattern formation, as pertinent to the individual dynamics associated with the two processes. Typical representative time evolution snapshots at $T=0.6T_c$ are presented in Fig.\ \ref{snapshots}, for different $p_r$. 
 For the highest $p_r=10^{-1}$, the patterns are similar to what is observed for a system with  purely nonconserved spin-flip dynamics, i.e., in \emph{phase ordering} \cite{majumder2023,Janke2023}. The \emph{isomerization} reaction seems to have finished faster as $p_r$ increases. This rationalizes the difference in the set of times for which the snapshots are presented for different $p_r$ in Fig.\ \ref{snapshots}. As $p_r$ decreases, the snapshots at intermediate times appear to have more bicontinuous  morphologies. For all cases, although at different times, eventually the system approaches a final morphology where the system has one of the \emph{isomers} as majority. For $p_r=10^{-4}$, such a stage is reached at a much longer time $\approx 5\times 10^7$ MCS, making it computationally expensive. Hence, we refrain ourselves from simulating larger lattices than $L=32$. Nevertheless, for investigating the reaction kinetics this size is sufficient as will be seen subsequently. For evolution snapshots at other temperatures see Figs.\ 1 and 2 in the Supporting Information (SI). From there it is apparent that at high temperature ($T=0.8T_c$) for smaller $p_r$ at first the system segregates to a slab-like morphology, and then evolves further due to the \emph{isomerization} reaction. However, at low temperature ($T=0.4T_c$) even for $p_r=10^{-4}$, the morphologies resemble more of what is shown in Fig.\ \ref{snapshots}.  
 
\begin{figure*}[t!]
\centering
\includegraphics*[width=0.45\textwidth]{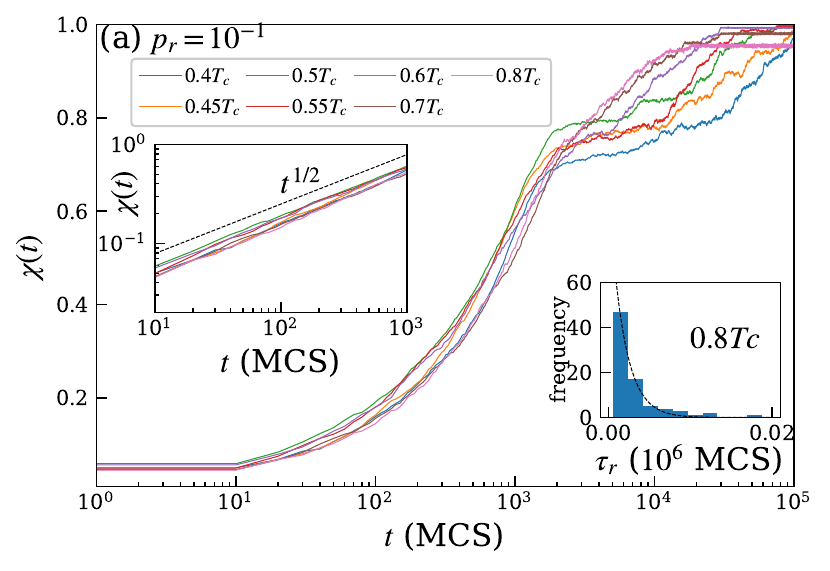}
\includegraphics*[width=0.45\textwidth]{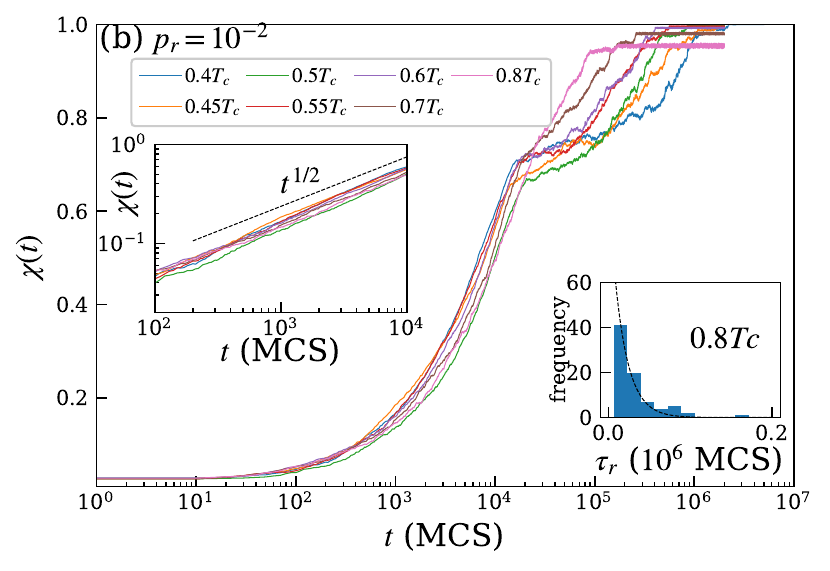}\\
\centering
\includegraphics*[width=0.45\textwidth]{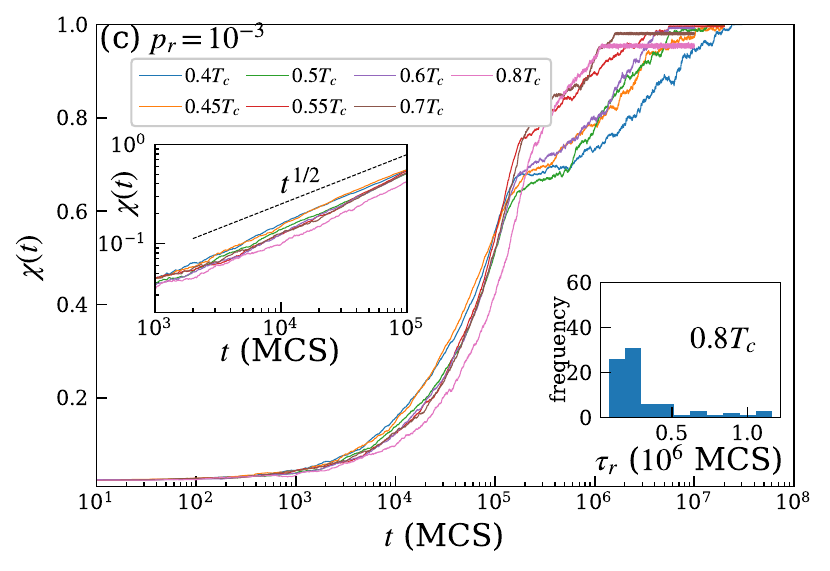}
\includegraphics*[width=0.45\textwidth]{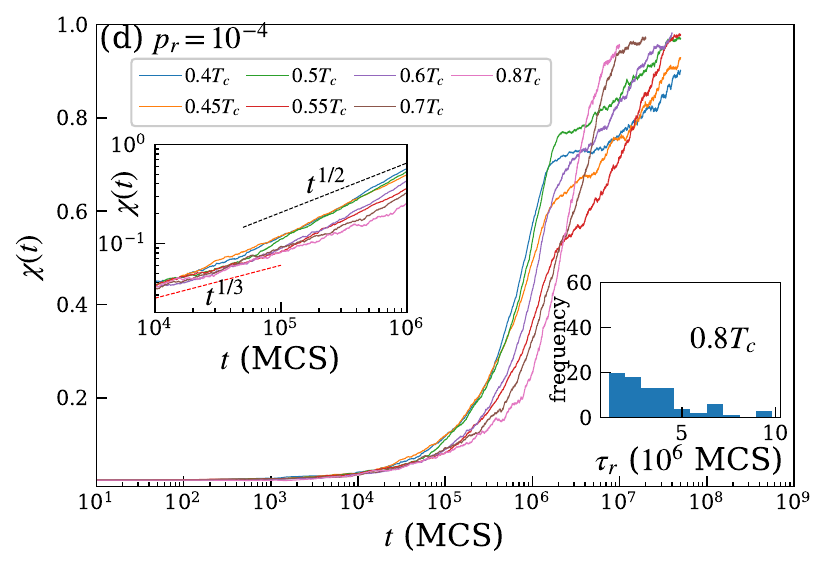}
\caption{\label{reac_prog} {\bf Progress of the reaction}. Linear-log plots for time dependence of the concentration difference  $\chi(t)$ of the two \emph{isomers}, at different temperatures for (a) $p_r=10^{-1}$, (b) $p_r=10^{-2}$, (c) $p_r=10^{-3}$, and (d) $p_r=10^{-4}$. The data presented are averaged over $80$ independent time evolutions obtained by using different random number seeds in the MC simulations. The upper insets show the same plots on double-log scale. There the dashed black lines represent a power law $\sim t^{1/2}$.  The red dashed line in (d) represents another power-law $\sim t^{1/3}$. The lower insets show representatives of histograms of the extracted reaction completion time $\tau_r$, at $T=0.8T_c$. In (a) and (b) the dashed lines represent a function of the form $f(\tau_r)=100\exp(-\lambda \tau_r)$, where $\lambda=550$ and $60$, respectively.}
\end{figure*}

\par
Since the main objective is to study the reaction kinetics, we have to extract the rate constant $k$. As a first step, we need to monitor the progress of the reaction until it finishes, i.e., when one of the molecular species becomes almost negligible compared to the other. For that we calculate the concentration difference of the two species 
\begin{equation}\label{conc_diff}
 \chi(t)=\frac{|N_{A_1}(t)-N_{A_2}(t)|}{N_{A_1}(t)+N_{A_2}(t)},
\end{equation}
where $N_{A_1}(t)$ and $N_{A_2}(t)$ are, respectively, the number of molecules of $A_1$ and $A_2$, at a time $t$. The denominator in Eq.\ \eqref{conc_diff}, $N_{A_1}(t)+N_{A_2}(t)=L^2$ is the total number of molecules present in the system. By construction, at $t=0$ for a \emph{racemic} mixture $\chi(0)=0$, and at large $t$ when the reaction is completed $\chi(t) \approx 1$, thus making $\chi(t)$ an useful parameter to monitor the progress of the reaction. In the main frames of Figs.\ \ref{reac_prog}(a)-(d), we present the corresponding data at different $T$ for four choices of $p_r$, as indicated. The linear-log scale is used to make the initial regimes properly visible. Apparently, for all $p_r$ the data show an initial transient regime when $\chi(t)$ remains almost constant, followed by a steep increase before finally settling at a value $\approx 1$ indicating completion of the reaction. However, one could notice that the transient regime broadens as $p_r$ decreases indicating a dominance of segregation over the reaction. Noticeable is also the presence of a third regime where $\chi(t)$ again attains almost a plateau before  finally approaching unity. For $p_r=10^{-4}$, at high $T$ this plateau vanishes (see Figs.\ 3-6 in the SI for individual plots at different $T$).
The upper insets of Figs.\ \ref{reac_prog} showing the same data on a double-log scale unravel significant differences in the time dependence of $\chi(t)$ for different $p_r$. For $p_r=10^{-1}$ and $10^{-2}$, data for all $T$ is consistent with a power-law $\chi(t) \sim t^{1/2}$. In a ferromagnetic system $\chi(t) \equiv |m(t)|$. There, during \emph{phase ordering} the absolute magnetization $|m(t)|$ obeys the same  power-law $|m(t)| \sim t^{1/2}$ in space dimension $d=2$ \cite{Janke2023}. For lower $p_r$, data for $\chi(t)$ deviate from the $\sim t^{1/2}$ behavior, particularly at high $T$.  This is owing to the dominance of segregation at low $p_r$, making the growth of $\chi(t)$ much slower with a power-law exponent of $1/3$, reminiscent of the Lifshitz-Slyozov exponent \cite{lifshitz1961} observed for the time dependence of the characteristic length during phase segregation \cite{majumder2010,majumder2011,majumder2018Potts}.
\begin{figure}[t!]
\centering
\includegraphics*[width=0.45\textwidth]{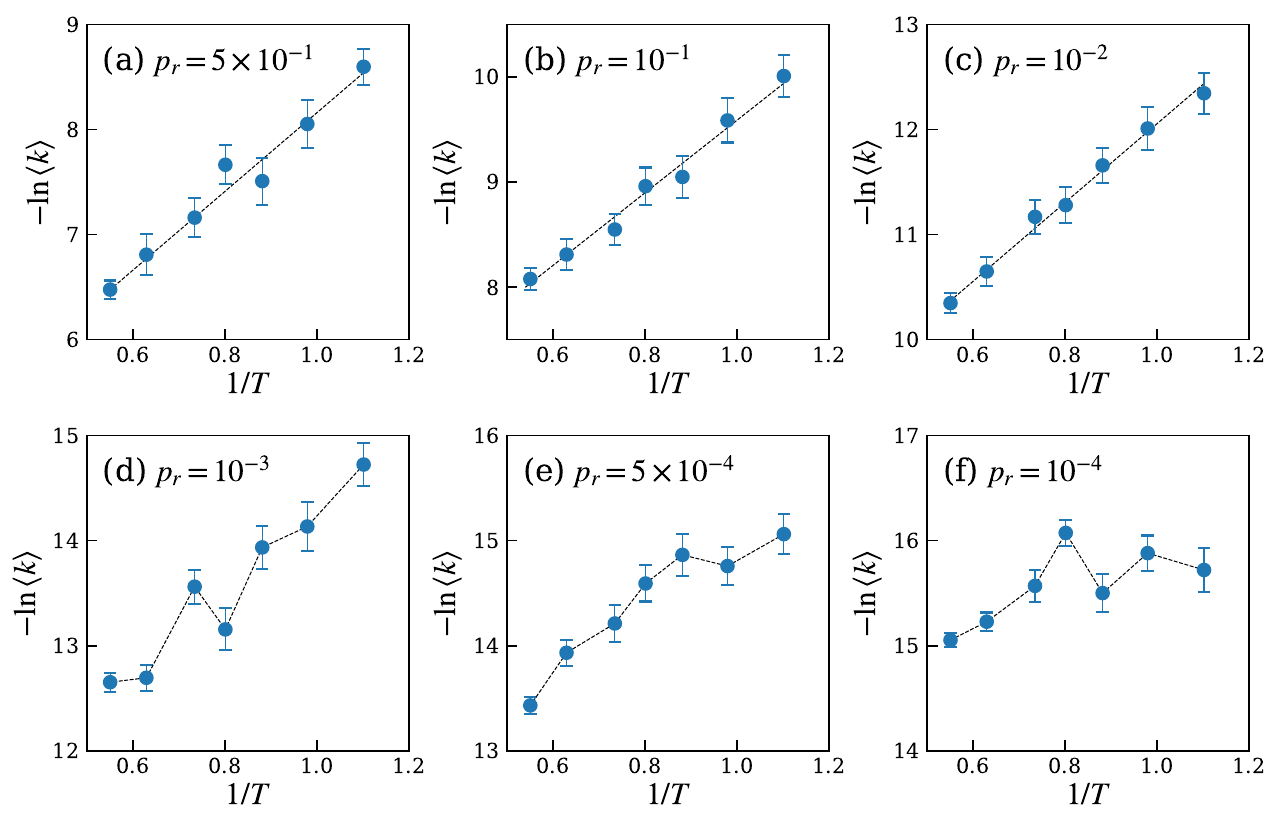}\\
\caption{\label{Arrhenius} {\bf Temperature dependence of the reaction rate constant} $\boldmath{k}$. Plots of $-\ln \langle k \rangle $ against $1/T$ to verify the presence of Arrhenius behavior for different $p_r$. The dashed straight lines in (a)-(c) are fits using Eq.\ \eqref{Arhenius_eqn}, and the dashed lines in (d)-(f) are just connecting the data points. Here, the symbol $\langle \dots \rangle$ indicates an average over $80$ independent simulation runs.}
\end{figure}

\par
Next, we extract the reaction-completion time $\tau_r$ from $\chi(t)$, as $\chi(t=\tau_r)=h$,
where we choose $h=0.9$ \footnote{The overall qualitative behavior of all the subsequent results are insensitive to any choice of $h>0.6$.}. Histogram of the extracted $\tau_r$ 
(see Figs.\ 7-10 in the SI for histograms at different $T$ for four values of $p_r$) shows non-uniform localized patterns for $p_r \ge 10^{-2}$ with an exponential behavior at high $T$ as presented in the lower insets of Figs.\ \ref{reac_prog}(a) and (b), respectively, for $p_r=10^{-1}$ and $10^{-2}$, at $T=0.8T_c$. The dashed lines there represent best fits using $f(\tau_r)=100\exp(-\lambda \tau_r)$, with decay constants $\lambda=550$ and $60$, respectively, for $p_r=10^{-1}$ and $10^{-2}$, implying a slower decay as $p_r$ decreases. For even lower $p_r$, the exponential nature is lost and the histogram appears to flatten out, as shown for $p_r=10^{-4}$ in the lower inset of Fig.\ \ref{reac_prog}(d).

\par
From the extracted $\tau_r$ we calculate the rate constant $k$ of the isomerization reaction as $k=\tau_r^{-1}$. In Figs.\ \ref{Arrhenius} we show the temperature dependence of $k$, by plotting $-\ln \langle k \rangle $ as a function of $1/T$. For $p_r\ge 10^{-2}$, the data show a linear nature confirming the Arrhenius behavior depicted in Eq.\ \eqref{Arhenius_eqn}. The dashed lines in Figs.\ \ref{Arrhenius}(a)-(c) represent respective best fits  obtained using the ansatz in Eq.\ \eqref{Arhenius_eqn}. The obtained activation energies are $E_a\in [3.46,3.75]$ with a mean of $\langle E_a\rangle =3.65(18)$. For all $p_r \ge 10^{-2}$, fits using $E_a=3.65$ in Eq.\ \eqref{Arhenius_eqn} also work reasonably well, indicating possibly a $p_r$-independent activation energy. For $p_r < 10^{-2}$, the data do not appear to be linear anymore, and in fact for $p_r=10^{-4}$ it becomes almost flat. This implies that the dominance of diffusive segregation dynamics disrupts the Arrhenius behavior of the \emph{isomerization} reaction, even though segregation itself is an Arrhenius process \footnote{See Fig.\ 11 in the SI to check how the segregation order parameter $\psi(t)$ captures the kinetics of a purely phase segregating system, and the corresponding Arrhenius behavior of the relaxation time $\tau_s$.}.

\par
To investigate the phenomenon of an interplay of two Arrhenius processes leading to a non-Arrhenius behavior, we probe the segregation using the time evolution of the order parameter 
\begin{equation}
 \psi(t)=\frac{1}{L^2}\sum_i \frac{|n_{A_1}^\square-n_{A_2}^\square|}{n_{A_1}^\square+n_{A_2}^\square},
\end{equation}
where the $\sum$ is over all lattice sites and $n_{A_1}^\square$ (or $n_{A_2}^\square$) is the
number of $A_1$ (or $A_2$) molecules in a sub lattice of size $\ell \times \ell$ with $\ell=1$ around a site $i$ of the parent square lattice. By construction $\psi$ for a segregated system is higher ($\approx 1$) than a homogeneous one. For a purely segregating system, the time when $\psi$ approaches $1$ provides a measure of the associated relaxation time $\tau_s$, and a plot of $-\ln\langle \tau_s^{-1}\rangle$ against $1/T$ confirms the  Arrhenius behavior (see Fig.\ 11 in the SI). 

\par
\begin{figure*}[t!]
\centering
\includegraphics*[width=0.45\textwidth]{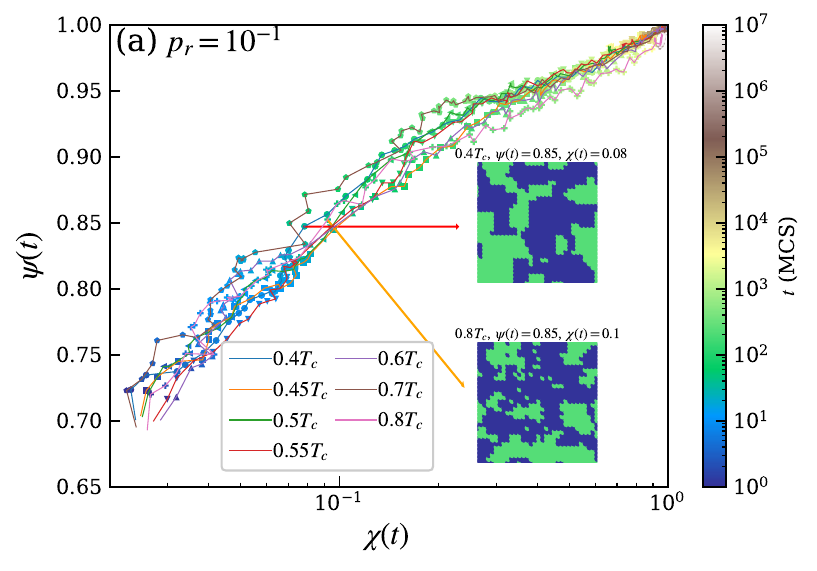}
\includegraphics*[width=0.45\textwidth]{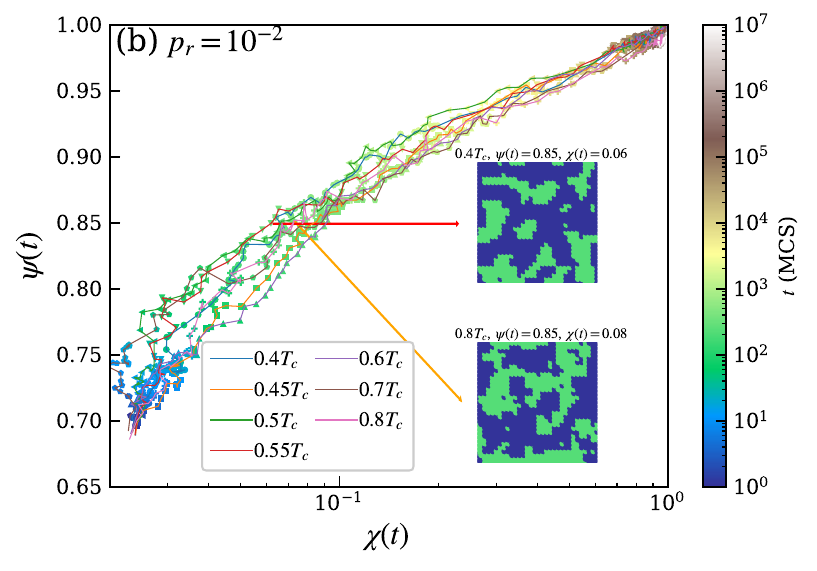}\\
\includegraphics*[width=0.45\textwidth]{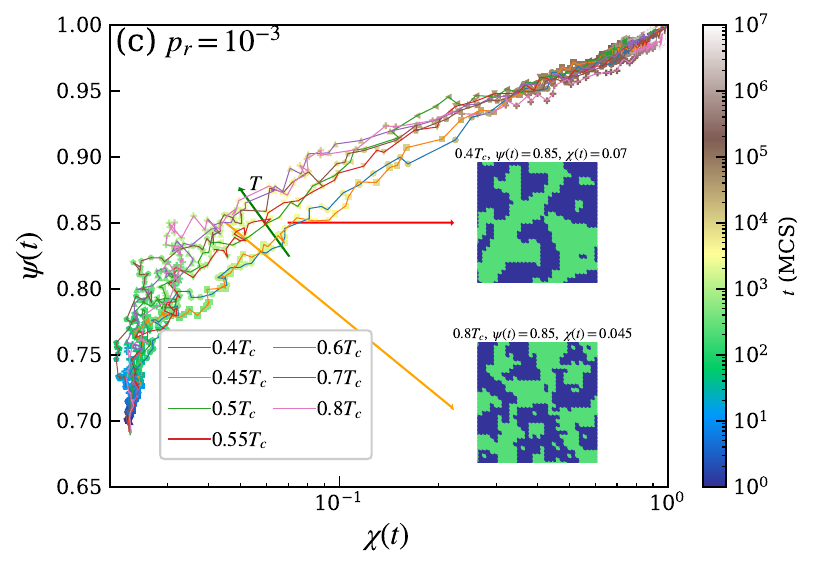}
\includegraphics*[width=0.45\textwidth]{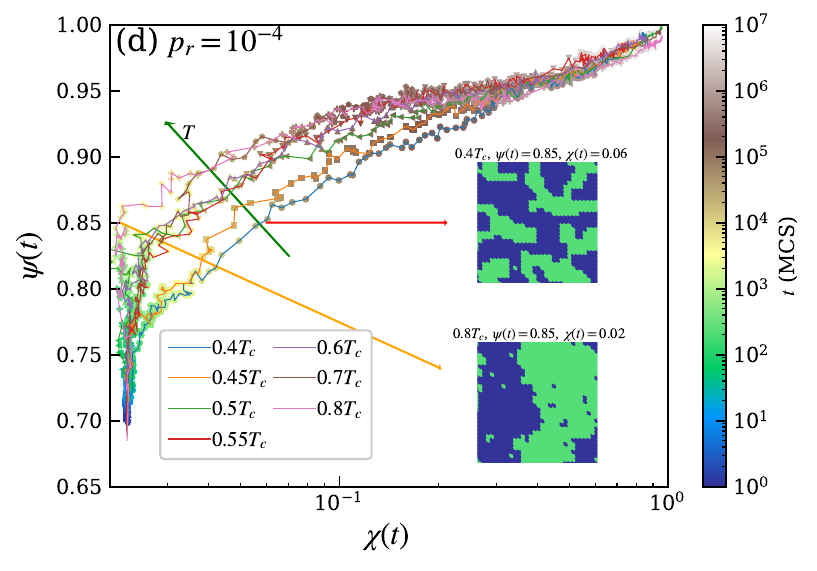}
\caption{\label{Arrhenius_reason} {\bf Interplay of segregation and reaction.} Plots of the time evolution of the segregation order parameter $\psi(t)$ against the corresponding concentration difference $\chi(t)$, on a linear-log scale at different temperatures $T$ for (a) $p_r=10^{-1}$, (b) $p_r=10^{-2}$, (c) $p_r=10^{-3}$, and (d) $p_r=10^{-4}$. The color bar represents the corresponding time. The snapshots at $T=0.4T_c$ and $0.8T_c$, represent typical configurations having $\psi(t)\approx 0.85$. The green arrows in (c) and (d) are guides to show the trend in data with increasing $T$.}
\end{figure*}

For a system where reaction is happening along with segregation, time dependence of $\psi(t)$ alone will capture the effect of both processes. Hence, to understand the interplay between the segregation and the reaction, in Figs.\ \ref{Arrhenius_reason} we plot the time dependence of $\psi(t)$, characterising the segregation, with $\chi(t)$, reflecting the temporal progress of the reaction, at different $T$ for four values of $p_r$. For $p_r = 10^{-1}$, shown in Fig.\ \ref{Arrhenius_reason}(a), $\psi(t)$ increases monotonously with $\chi(t)$, and the spread of data points over time for different $T$ appear to be quite condensed, suggesting no trend as a function of $T$.  Typical configurations having $\psi(t)\approx 0.85$, a value that corresponds to an almost completely segregated state for a purely segregating system, are also shown in Fig.\ \ref{Arrhenius_reason}(a) for a high and low $T$. None of them represent a completely segregated morphology, suggesting that both the dynamics affect the system concurrently. However, the reaction has progressed slightly further for $T=0.8T_c$ with $\chi(t)=0.1$ compared to $\chi(t)=0.08$ at $T=0.4T_c$. This difference eventually gets manifested in the form of an Arrhenius behavior, expected for a simple \emph{isomerization} reaction. 
\par
As $p_r$ decreases the data for different $T$ look more dispersed with a certain $T$ dependent trend, as guided by the green arrows in Figs.\ \ref{Arrhenius_reason}(c) and (d).  There, one can notice that at the beginning $\psi(t)$ increases sharply while no significant change in $\chi(t)$ is observed, implying that initially during the evolution segregation dynamics dominate. The effect is more pronounced for $p_r=10^{-4}$ at high $T$. This could be further appreciated from the almost completely segregated morphology of the configuration representing a system having $\psi(t)=0.85$ at $T=0.8T_c$, for $p_r=10^{-4}$,  shown in Fig.\ \ref{Arrhenius_reason}(d). The value of $\chi(t)=0.02$ at this instance indicates that the progress of the reaction is negligible. Since segregation itself is an Arrhenius process, the system reaches such a state much faster at high $T$. However, after attaining such a morphology not only the segregation dynamics almost seizes, the activation energy $E_a$ of the reaction also increases, temporarily halting the entire evolution of the system. This is analogous to the phenomenon of dynamic freezing due to emergence of metastable slab-like configurations during \emph{phase ordering} of a ferromagnet \cite{Vinals1986,olejarz2013,majumder2023}. On the other hand, at $T=0.4T_c$ the value of $\chi(t)=0.06$ when $\psi(t)=0.85$, suggests that the reaction has progressed further compared to $T=0.8T_c$. Corresponding typical configuration at $T=0.4T_c$, shown in Fig.\ \ref{Arrhenius_reason}(d), also does not represent a segregated morphology. Thus, in this case the reaction can easily proceed even further toward its completion. Hence, at low $T$ and low $p_r$, although the system always encounter a simultaneous occurrence of segregation and reaction, it never gets trapped in a completely segregated state, making the reaction completion time comparable with the one at high $T$. Overall it implies that for low $p_r$, the activation energy $E_a$ of the \emph{isomerization} reaction is not $T$ independent, and rather it depends irregularly on $T$, which in turn gets manifested in the form of a non-Arrhenius behavior of the \emph{isomerization} reaction. 
\par
To summarize, we have presented results on chemical kinetics of a simple \emph{isomerization} reaction when it competes with a segregation process among the \emph{isomers}. Results from our MC simulations of a model constructed using the nearest neighbor Ising model with two competing dynamics, reveal that the Arrhenius behavior of the reaction gets disrupted as segregation dynamics dominate over the reaction dynamics, even though segregation itself is an Arrhenius process. We have rationalized this observation by virtue of a phenomenological argument that at high temperature and low reaction probability, the  segregation of \emph{isomers} reaches completion leading to an almost completely segregated morphology, and thereby raising the activation energy of the reaction making it difficult for the system to evolve further. These findings shall provoke
an experimental verification, which to the best of our understanding can be
done without much hassle.
\par
As a next step it would be worth exploring aging and related dynamical scaling \cite{henkel_book} in a system with mixed dynamics such as presented here. Note that the results presented here are for solid phase reactions. Thus, as a future endeavor, it would be intriguing to consider similar reactions in solution phase by performing MD simulations of a fluid system \cite{MajumderPRL}. Furthermore, based on the model presented here, one can construct similar models for reactions of higher complexity, and subsequently may also invoke the role of a catalyst. For that use of a multi-species model like the Potts model is required \cite{majumder2023}.  

\acknowledgments
This work was funded by the Science and Engineering Research Board (SERB), Govt.\ of India for a Ramanujan Fellowship (file no.\ RJF/2021/000044). 

% \bibliography{bib.bib}
%Control: production of eprint (-1) disabled
%
\pagebreak

\begin{center}
\textbf{\large Supporting Information: Segregation Disrupts the Arrhenius Behavior of an Isomerization Reaction}
\end{center}
%%%%%%%%%% Merge with supplemental materials %%%%%%%%%%
%%%%%%%%%% Prefix a "S" to all equations, figures, tables and reset the counter %%%%%%%%%%
\setcounter{equation}{0}
\setcounter{figure}{0}
\setcounter{table}{0}
\setcounter{page}{1}
\makeatletter
\renewcommand{\theequation}{S\arabic{equation}}
\renewcommand{\thefigure}{S\arabic{figure}}
\renewcommand{\bibnumfmt}[1]{[S#1]}
\renewcommand{\citenumfont}[1]{S#1}
Here we present supporting information that includes figures we left out in the main
manuscript for brevity. It contains

\begin{itemize}
  \item Evolution snapshots at other $T$.
  \item Individual plots for progress of the reaction at different $T$. 
  \item Histograms of $\tau_r$ at different $T$ and $p_r$. 
  \item Time evolution of $\psi(t)$ for a purely segregating system and its corresponding Arrhenius behavior.
\end{itemize}
\begin{figure*}[b!]
	\centering
	\includegraphics*[width=0.5\textwidth]{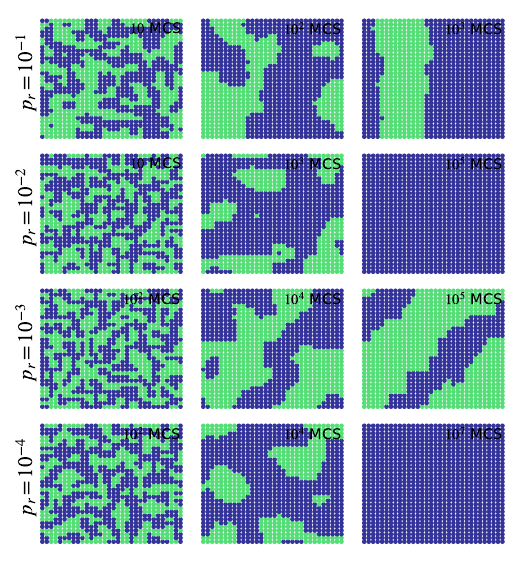}\\
	\caption{ Time evolution snapshots at a relatively low temperature $T=0.4T_c$ for different $p_r$.}
\end{figure*}
\begin{figure*}[t!]
	\centering
	\includegraphics*[width=0.5\textwidth]{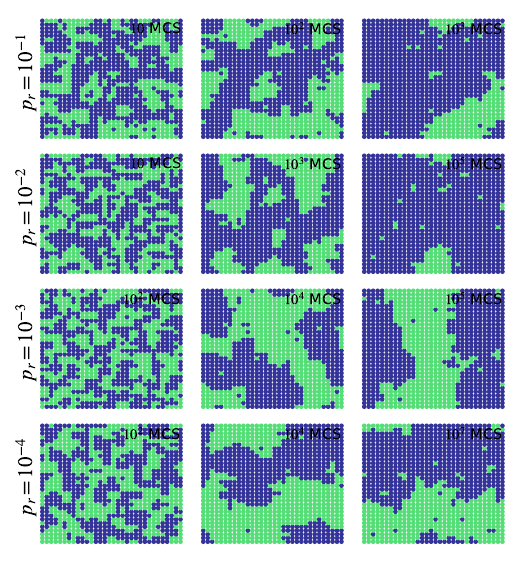}\\
	\caption{ Time evolution snapshots at a relatively high temperature $T=0.8T_c$ for different $p_r$.}
\end{figure*}

\begin{figure*}[t!]
	\centering
	\includegraphics*[width=0.45\textwidth]{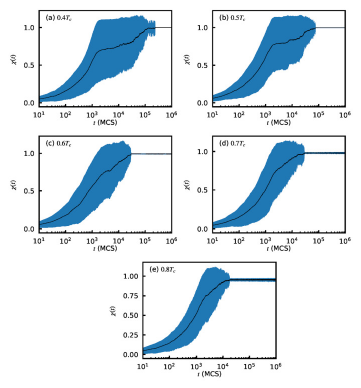}\\
	\caption{ Progress of the reaction in terms of the concentration difference of the two molecular species $\chi(t)$, at different temperatures for $p_r=10^{-1}$. The black solid lines represent averaging over $80$ independent simulation runs while the shaded regions indicate the corresponding standard deviations.}
\end{figure*}
\begin{figure*}[b!]
	\centering
	\includegraphics*[width=0.45\textwidth]{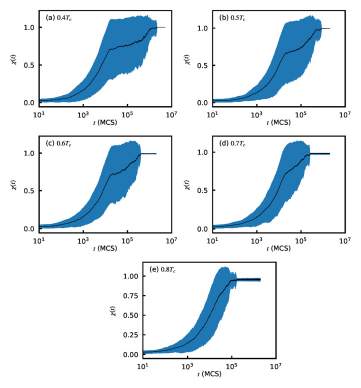}\\
	\caption{ Progress of the reaction in terms of the concentration difference of the two molecular species $\chi(t)$, at different temperatures for $p_r=10^{-2}$. The black solid lines represent averaging over $80$ independent simulation runs while the shaded regions indicate the corresponding standard deviations.}
\end{figure*}
\begin{figure*}[t!]
	\centering
	\includegraphics*[width=0.45\textwidth]{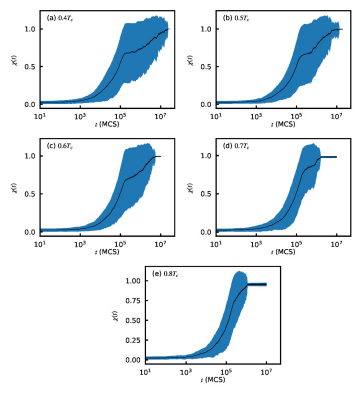}\\
	\caption{ Progress of the reaction in terms of the concentration difference of the two molecular species $\chi(t)$ at different temperatures for $p_r=10^{-3}$. The black solid lines represent an average over $80$ independent simulation runs while the shaded regions indicate the corresponding standard deviations.}
\end{figure*}
\begin{figure*}[b!]
	\centering
	\includegraphics*[width=0.45\textwidth]{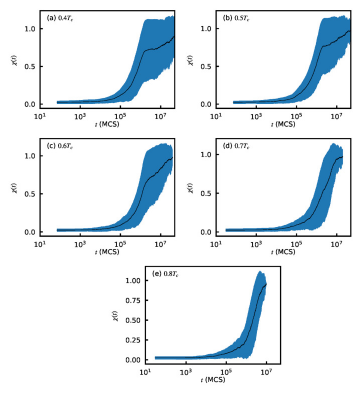}\\
	\caption{ Progress of the reaction in terms of the concentration difference of the two molecular species $\chi(t)$ at different temperatures for $p_r=10^{-4}$. The black solid lines represent an average over $80$ independent simulation runs while the shaded regions indicate the corresponding standard deviations.}
\end{figure*}
\begin{figure*}[t!]
	\centering
	\includegraphics*[width=0.45\textwidth]{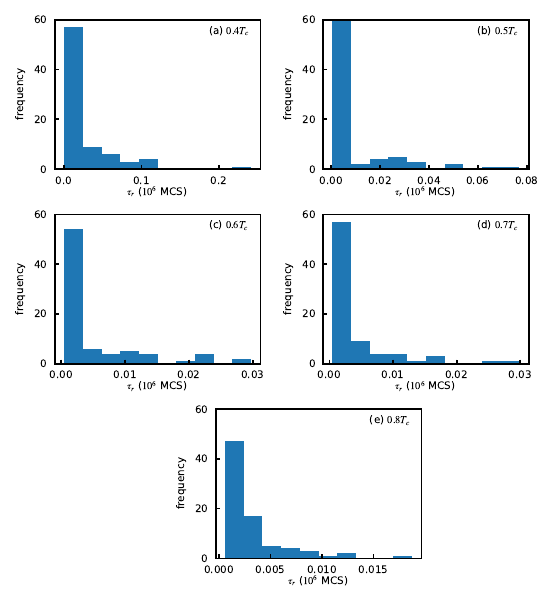}\\
	\caption{ Histograms of the reaction-completion time $\tau_r$ at different temperatures for $p_r=10^{-1}$.}
\end{figure*}

\begin{figure*}[b!]
	\centering
	\includegraphics*[width=0.45\textwidth]{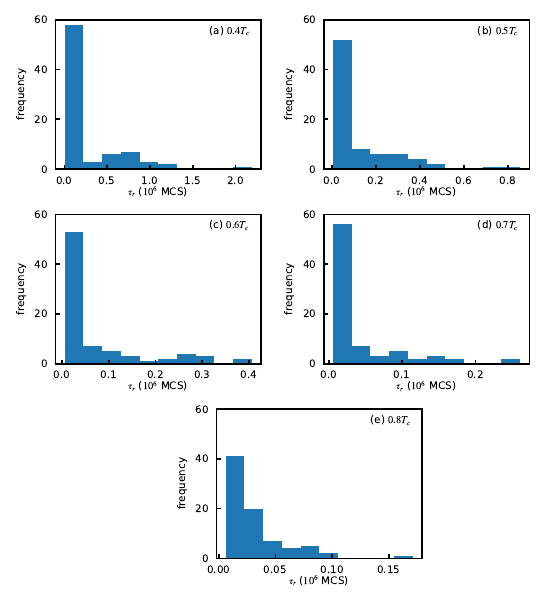}\\
	\caption{ Histograms of the reaction-completion time $\tau_r$ at different temperatures for $p_r=10^{-2}$.}
\end{figure*}
\begin{figure*}[t!]
	\centering
	\includegraphics*[width=0.45\textwidth]{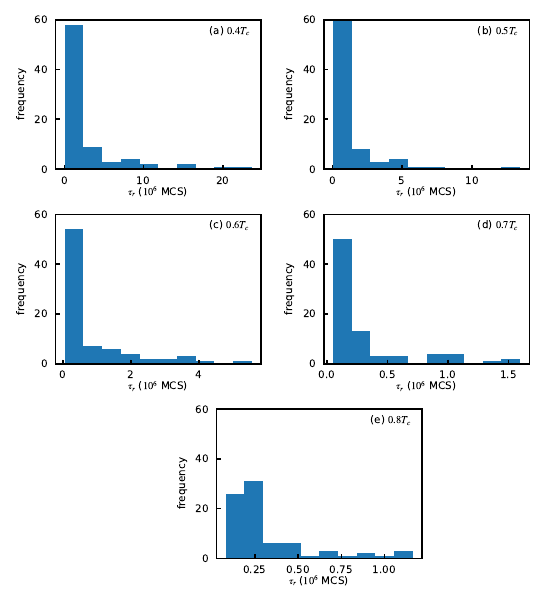}\\
	\caption{ Histograms of the reaction-completion time $\tau_r$ at different temperatures for $p_r=10^{-3}$.}
\end{figure*}
\begin{figure*}[b!]
	\centering
	\includegraphics*[width=0.45\textwidth]{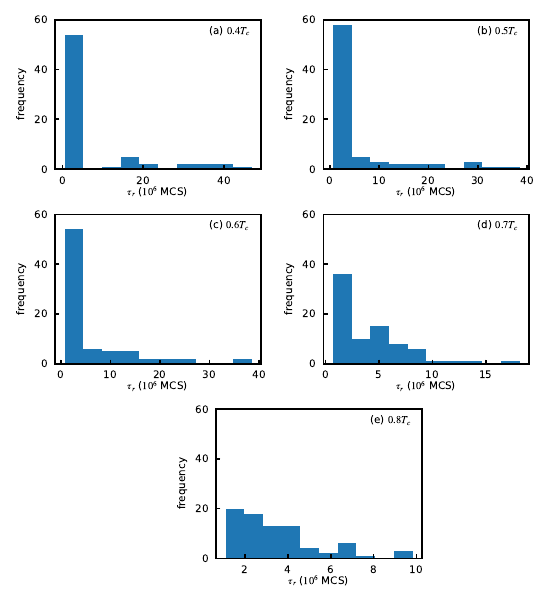}\\
	\caption{ Histograms of the reaction-completion time $\tau_r$ at different temperatures for $p_r=10^{-4}$.}
\end{figure*}
\begin{figure*}[t!]
	\centering
\includegraphics*[width=0.5\textwidth]{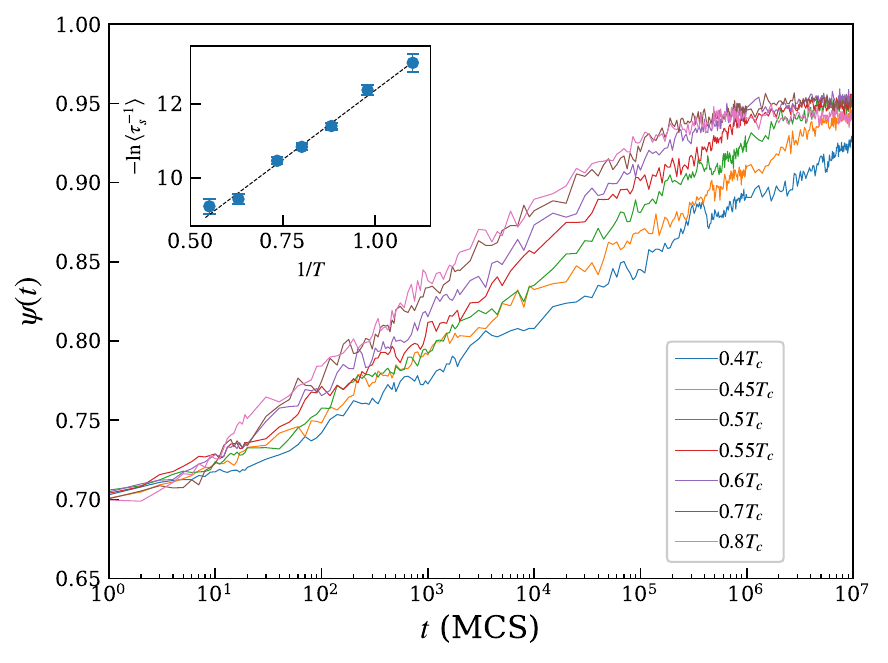}\\
	\caption{Time evolution of the segregation order parameter for a purely segregating system at different $T$. The inset shows the Arrhenius behavior of the extracted relaxation $\tau_s$ for the segregation in terms of a plot of $-\ln\langle \tau_s^{-1}\rangle$ against the inverse temperature $1/T$. The dotted line is the best fit line representing Eq.\ (2) from the main text.}
\end{figure*}
\end{document}